\documentclass[twocolumn,showpacs,preprintnumbers,amsmath,amssymb]{revtex4}
\usepackage{graphicx}
\input epsf
\usepackage{dcolumn}
\usepackage{bm}

\begin{document}


\title{Can we escape from the big rip
in the achronal cosmic future?}

\author{A.V. Yurov}
\email{artyom_yurov@mail.ru}
\author{V.A. Yurov}%
 \email{yurov@freemail.ru}
\author{S. D. Vereshchagin}%
 \email{sergev@nightmail.ru}
\affiliation{%
The Theoretical Physics Department, Kaliningrad State
University,A. Nevskogo str., 14, 236041,
 Russia.
\\
}%


\date{\today}

\begin{abstract}
Recently Pedro F. Gonz\'alez-D\'iaz had shown that phantom energy
can results in achronal cosmic future where the wormholes become
infinite before the occurrence of the big rip singularity. After
that, while the phantom energy continues its accretion, any
wormhole becomes the Einstein-Rosen bridge. Pedro F.
Gonz\'alez-D\'iaz has suggested that such objects can be used by
an advanced civilization as the means of escape from the big rip.
Using the Bekenstein Bound we'll show that it is impossible due to
the very strong upper bound laid on the total amount of
information which can be sent through this bridge.
\end{abstract}

\pacs{98.80.Hw, 04.70.-s}
\maketitle


The cosmology nowadays is amazingly abundant with a new startling
solutions. Some of the most recent ones are the models with the
"Phantom fields" which result in the violation of the weak energy
condition (WEC) $\rho>0$, $\rho+p/c^2>0$ \cite{1}, \cite{2}, where
$\rho$ is the fluid density and $p$ is the pressure. Such phantom
fields, as follows from their quantum theory \cite{3}, should
inevitably be described by the scalar field with the negative
kinetic term. The through investigations shows that such fields
are apparently could not be considered as a fundamental objects.
However, it is possible that the Lagrangians with the negative
kinetic terms will appear as some kind of effective models, as it
happens in some models of supergravity \cite{4} and in the gravity
theories with highest derivatives \cite{5}. Finally, the "phantom
energy" in the brane theory was considered in \cite{7}.

The particular interest to models with the phantom fields is
caused by their prediction of so-called "Cosmic Doomsday" alias
big rip \cite{1} (see also \cite{8}). In case of the phantom
energy we have $w=p/(c^2\rho)=-1-\epsilon$ with $\epsilon>0$.
Integration of the Einstein-Friedmann equation for the flat
universe results in
\begin{equation}
\begin{array}{cc}
\displaystyle{
a(t)=\frac{a_0}{\left(1-\xi t\right)^{2/3\epsilon}},}\\
\displaystyle{
\rho(t)=\rho_0\left(\frac{a(t)}{a_0}\right)^{3\epsilon}=\frac{\rho_0}{(1-\xi
t)^2}}, \label{1}
\end{array}
\end{equation}
where $\xi=\epsilon\sqrt{6\pi G\rho_0}$. We choose $t=0$ as the
present time, $a_0\sim 10^{28}$ cm and $\rho_0$ to be the present
values of the scale factor and the density. There, if
$t=t_*=1/\xi$, we automatically get the big rip.

Recently Pedro F. Gonz\'alez-D\'iaz had shown that phantom energy
can results in achronal cosmic future where the wormholes become
infinite before the occurrence of the big rip singularity. To show
this lets consider the wormhole with the throat radius
$b_0=10^{-33}$ cm (Planck scale). It was shown in \cite{9} that if
$p =-(1+\epsilon)c^2\rho$ is a fluid's equation of state, then
\begin{equation}
c{\dot b}(t)=2\pi^2\epsilon GD\rho(t)b^2(t), \label{dotb}
\end{equation}
where $b(t)$ is the throat radius of a Morris-Thorne wormhole and
$D$ is dimensionless quantity. According to \cite{9} we can choose
$D\sim 4$ (see also \cite{10}). The equation (\ref{dotb})
describes the changing of the $b(t)$ with regard to the phantom
energy's accretion. Integration of the (\ref{dotb}) gets us
\begin{equation}
\frac{1}{b(t)}=\frac{1}{b_0}-\frac{2\pi^2\epsilon\rho_0 GD
t}{c(1-\xi t)}. \label{1b}
\end{equation}
Therefore at
\begin{equation}
{\tilde t}=\frac{c}{\epsilon(c\sqrt{6\pi G\rho_0}+2\pi^2\rho_0b_0G
D)} \label{tild}
\end{equation}
we get $b({\tilde t})=\infty$. As we can see ${\tilde t}<t_*$, and
therefore this universe indeed will be achronal before the
occurance of the big rip. In accord to \cite{9}, at $t>{\tilde
t}$, while in process of the phantom energy's accretion, the
wormhole becomes an Einstein-Rosen bridge which (again according
to \cite{9}) can, in principle, be used by the future advanced
civilizations in their efforts to escape from the big rip.
However, in this activity the future civilizations will have to
face but another problem: an extremely stringent information's
bound.

Indeed, the Bekenstein Bound \cite{11} shows that the total amount
of information, which can be stored in region of radius $R$ is
$I<I_{m}=2\pi RE/(\hbar c\log 2)$ thus
\begin{equation}
I<I_{m}=2.5786\times 10^{43}\,\left(\frac{M}{1\,\,{\rm
kilogram}}\right)\left(\frac{R}{1\,\,{\rm meter}}\right)\,\,{\rm
bits}. \label{BB}
\end{equation}
Substituting $\rho_0=\rho_c=10^{-29}$ ${\rm gramme}/{\rm cm}^3$,
$R=c/H_0$, and $H_0\sim 70$ km/s/Mps (the current measured value
of the Hubble constant) results in
\begin{equation}
 I_m=0.33\times 10^{123}\,\,{\rm bits}
 \label{naiv}.
\end{equation}

Recently, Krauss and Starkman \cite{12}, upon the usage of more
detailed calculations have shown that the currently observed
acceleration of the Universe can induce the universal limit on the
total amount of information that can in principle be stored and
processed in the future. The maximal amount of information has
appeared to be $I_{_{max}}\sim 1.35\times 10^{120}$\cite{Moore}.
Our naive result (\ref{naiv}) is consistent with the detailed
calculations in \cite{12}. Due to this reason, we will further use
the crude estimation (\ref{BB}) instead of detailed calculations
like in \cite{12}.

In general case, the horizon distance can be calculated as
\begin{equation}
R_c(t)=a(t)\int_t^{t_{_{max}}} \frac{c dt'}{a(t')}. \label{Rc}
\end{equation}
Substituting (\ref{1}) into the (\ref{Rc}) gives
\begin{equation}
R_c(t)=\frac{3c\epsilon(1-\xi t)}{(2+3\epsilon)\xi}, \label{Rcph}
\end{equation}
for the case of phantom field ($t_{_{max}}=t_*$). This, of course,
means that $R_c(t)\to 0$ as $t\to t_*$.

Substituting (\ref{tild}) into the (\ref{Rcph}) we get
\begin{equation}
R_c({\tilde t})=\frac{\pi^2\rho_0G Db_0c\sqrt{6}}{(\sqrt{6\pi
G\rho_0c^2}+2\pi^2\rho_0GDb_0)\sqrt{\pi G\rho_0}(2+3\epsilon)}.
\label{Rtild}
\end{equation}
Since $\epsilon\ll 1$ the value of $R_c({\tilde t})$ is $\sim
0.6\times 10^{-32}$ cm. Thus the horizon's radius has the order of
the Planck scale. Finally, the Bekenstein Bound (\ref{BB}) results
in
\begin{equation}
I<I_m({\tilde t})\sim \frac{2.74 R^4_c({\tilde
t})\rho_0}{\left(1-\xi{\tilde t}\right)^2}\times 10^{43}\sim
68.66\,\,{\rm bits}. \label{Imm}
\end{equation}

Thus, the largest amount of information to be processed and
therefore - sent through the bridge is no way greater then 57
bits. As a comparison, the superior limit of amount of information
encoded in a human being is about $10^{45}$ bits. This value is
surely an excess, for any object existing in current universe is
encoded far less than any quantum field theory's constraints, but
in any case the amount of 69 bits seems terribly small. For
example, the amount of information in a typical book is $I\sim
10^{7}$ bits, and $10^{15}$ bits stands for all books in the
Library of Congress (we adopt this example from the work
\cite{15}). 69 bits is, in fact, slightly higher then the total
amount of information that can be coded in the proton, viz. 44
bits! (see \cite{13})

Our conclusion is founded on the assumption of the big rip's
existence. But the big rip's appearance can be prevented via some
certain ways. For example, in \cite{14} it is argued that the
realistic cosmological evolution should include the quantum
effects, and their accounting results in universe, ending up in de
Sitter phase before the scale factor $a(t)$ blows up. From this
point of view, our calculations should be revised in the following
way: in order to use the wormholes as the gate to another domains
one has to wait till the moment ${\tilde t}$, i.e. the time when
the wormholes become infinite. The whole scenario should be
following: at $t<{\tilde t}$ the phantom energy dominates and one
gets an infinite wormhole that is transformed into the
Einstein-Rosen bridge. Then, at $t>{\tilde t}$ the quantum effects
will dominate, resulting in the quantum escape from the big rip.
Approximately, at $t<{\tilde t}$ we have the ''phantom universe''
whereas at $t>{\tilde t}$ we have the de Sitter universe. During
de Sitter phase the throat radius of a Morris-Thorne wormhole is
constant and the same is true for the Hubble expansion rate
$H_{_{dS}}$. So, roughly speaking, one can assume that
$H_{_{dS}}=H({\tilde t})$ (in fact, $H_{_{dS}}>H({\tilde t})$ but
this can only strengthen our conclusion). Therefore, the usage of
the Bekenstein Bound requires the substitution of
$R=c/H_{_{dS}}=c/H({\tilde t})$ into the (\ref{BB}), with
$$
H({\tilde t})=\frac{2\xi}{3\epsilon(1-\xi{\tilde t})},
$$
and with ${\tilde t}$ taken from the (\ref{tild}). The simple
calculations results in
$$
I<I_m=68.66\,\,\,{\rm bits},
$$
that is sufficiently the same as (\ref{Imm}).

Another interesting probability - the possible escape from the our
domain (without the phantom fields) was considered in \cite{15}.
In this work authors have proposed to use the act of formation of
a new inflating regions. They have shown that the probability of
the nucleation will be largest if the energy scale of inflation is
close to the Planck scale. However, the information's bound exists
even in this case: the sufficiently large amounts of information
that should be sent necessarily requires the smallness of the
Hubble root $H$, which, in turn, greatly reduces the very
probability of such nucleation. For example, the amount of
information contained in human genome results in tunneling
probability $P\sim \exp\left(-10^{11}\right)$, whereas for the
information from all books of the Library of Congress the
probability is $P\sim \exp\left(-10^{16}\right)$.

As we have seen, the way of Pedro F. Gonz\'alez-D\'iaz results, as
well, in information limitation which is even more hardly then the
information bound from the \cite{15}. Therefore, it is very
problematic even for advanced civilization to use this scenario in
order to escape from the big rip or even to leave our domain. And
what is more, even if the results of work \cite{14} are correct
and phantom universe will transform into the de Sitter phase,
(i.e. into the phase of new inflation) due to the quantum effects,
and even if advanced civilization will decide to stay in the same
domain, they absolutely have no chances to go through the moment
$t={\tilde t}$ to continue it's existence in new universe. Nothing
more complex then proton will.
\newline
\newline


\end{document}